\documentclass[5p,10pt]{elsarticle}
\usepackage{amsmath}
\usepackage{amssymb}
\usepackage{graphicx}
\usepackage{psfrag}
\usepackage{wasysym}
\usepackage{color}
\usepackage{mathrsfs}

\usepackage{amssymb}

\newcommand{\id}{\mathrm{d}} 

\journal{arXiv}

\begin{document}

\begin{frontmatter}



\title{Sensing and Multiscale Structure}
\author{John F.~A.~Fletcher\corref{cor1}}
\ead{j.f.a.fletcher@damtp.cam.ac.uk}
\cortext[cor1]{Corresponding author. Postal address: DAMTP, Centre for Mathematical
  Sciences, Wilberforce Road, Cambridge CB3 0WA. Telephone: +44 1223
  765000. Fax: +44 1223 765900.} 
\address{DAMTP, University of Cambridge, Wilberforce Rd,
  Cambridge, CB3 0WA, UK}


\begin{abstract}
We introduce a method of estimating parameters associated with a fractal
 random scattering medium, which utilizes the multiscale properties of the
 scattered field. The example of ray-density fluctuations beyond a
 phase screen with fractal slope is considered. An exact
 solution to the forward problem, in the case of the Brownian fractal,
 leads to an expression for the volatility of the slope. This
 expression is invariant under a change of probability measure, a fact
 which gives rise to the 
corresponding result for a (stationary) Ornstein-Uhlenbeck slope. We
demonstrate that our
analytical results are consistent with numerical simulations. Finally,
an application to the determination of sea ice thickness via sonar is discussed.
\end{abstract}

\begin{keyword}
Inverse scattering \sep Parameter estimation \sep Brownian motion
\end{keyword}

\end{frontmatter}


\section{Introduction}
\label{Introduction}

When a wave is scattered by a random medium the 
scattered radiation can be described by a random field. We consider
the problem of estimating parameters related to the characteristics of
the scattering medium. The numerous 
applications of such problems, for example, in the areas of industry,
geophysics and medicine are well known, and are of
considerable practical importance \cite{colton00}.

One possible approach is the following: First, 
derive an expression for the ``random scattered field'', or some
observable properties thereof. Then, compare these theoretical quantities with
the corresponding sample quantities derived from experiment. A popular 
version of this strategy is the ``method of moments'' which uses 
the mean, variance, correlation function etc. A refinement of this,
also along Bayesian lines, is the method of maximum likelihood, which can provide
optimal estimates of the parameters in question \cite{Bishwal08}. However, this technique is not usually available in
scattering problems. Elementary considerations
soon lead to the conclusion that scattered fields (arising from, for
example, a homogeneous random scattering medium) are rarely endowed with convenient properties. For instance, in the example of surface scattering
the Markov property will not hold in general, since the scatter could
be a function of the whole of the scattering surface. 

If the scattered field has fractal properties,
there is another possibility for estimating parameters, unrelated to
Bayes theorem. Information can be derived from 
fluctuations in the field at all scales. The principle
can be understood by relation to the ``well known fact'' that the volatility 
of an Ito diffusion can be recovered with probability one, given a sample path
over any non-zero interval, see e.g. \cite{Rogers00}. It 
is common for naturally occurring structures to possess multiscale character
suitable to be modeled as a fractal. Moreover, it seems reasonable 
to suppose that such structures might confer
multiscale properties to the scattered field. Parameter estimation based around this idea
is potentially very efficient, since it does not require the availability 
of data over many correlation lengths. This approach to
sensing (inverse scattering) is apparently new,
and is illustrated here through the example of ray-density
fluctuations beyond a subfractal phase-changing screen (SPS). The concept of
rays is of great utility in the theory of scattering, most famously in
the shortwave limit \cite{keller}, but also
more generally \cite{Budaev07}. The
ray-density, denoted by $R$, describes
intensity fluctuations induced by a SPS in an incoherent configuration (no
interference). $R$ also features in expressions for the
moments of the scattered intensity
in a coherent configuration \cite{jakeman82}.
We consider scatter from a one-dimensional SPS in an incoherent configuration,
which provides the simplest example of the proposed technique.
\section{The phase screen model}
\label{ray}
\begin{figure}[h]
\psfrag{A}[c]{{\small $a$}}
\psfrag{B}[c]{{\small $\sigma$}}
 \includegraphics[width=1.0\linewidth]{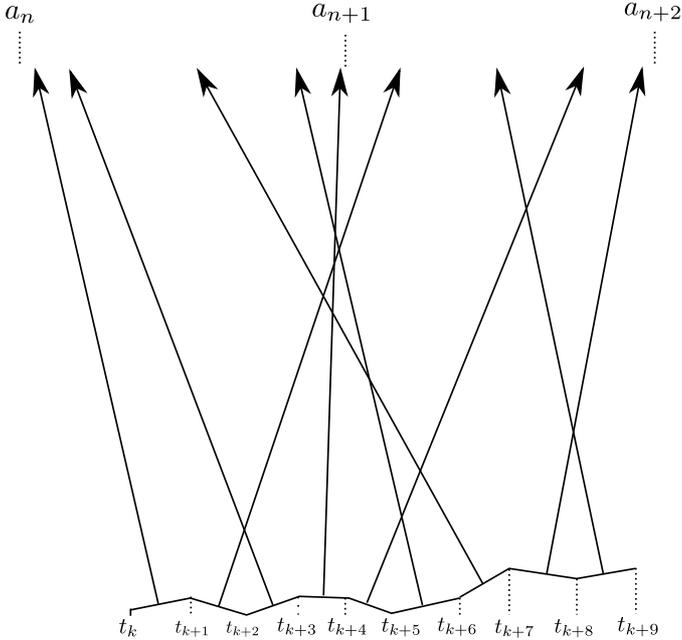}
\caption{Rays scattered from a faceted random surface, or through a
  subfractal phase-changing screen.}
\label{fig1}
\end{figure}

When a plane of parallel rays passes through a non-flat refracting layer, differences in the refractive index of the
layer cause the rays to scatter. The situation may be described by specifying
the slope (gradient) of a surface of constant phase in terms of a one-dimensional stochastic process
$(Y_t)_{t\in \mathbb R}$. 
If $Y$ is
fractal, then we have a ``subfractal
phase-changing screen'' \cite{jakeman82}. This model also describes surface scattering,
when the source and
point of observation are coincident, and shadowing and multiple
scattering are neglected, see Fig. \ref{fig1}. These surfaces
can be thought of as possessing a faceted structure, and have been used to model, among other things,
scattering of radio waves in the ionosphere \cite{rino79}, and the effect of 
internal waves on acoustic propagation in the ocean \cite{uscinski81}. 

The ray-density can be approximated at $a\in A_n:=[a_n,a_{n+1})$
where $a_n=n\Delta a$, a unit distance from
the layer by $\{\mathbf R_n^Y(\Delta t,\Delta a);n\in \mathbb Z\}$
defined as

\begin{equation}\label{eq:nrw1}
\mathbf R_{n}^Y(\Delta t,\Delta a):=\frac{1}{\Delta a}\sum_{k=-\infty}^\infty
\mathbf{1}_{\{t_k-{Y}_{t_k}\in A_n\}}\Delta t
\end{equation}
where $t_k=k\Delta t$ and $\Delta a \gg \Delta t$. The
ray-density $(R_a^Y)_{a\in\mathbb{R}}$ is defined as the limit (when it exists)
\begin{equation}\label{eq:One}
R_a^Y:=\lim_{\Delta a\rightarrow 0}\lim_{\Delta t\rightarrow 0}\mathbf
R_{n}^Y(\Delta t,\Delta a)=:\int_{\mathbb{R}} \id t\,\delta(Y_t-t+a)
\end{equation}
where $a\in A_n$. Assume that $Y$ is an \emph{Ito diffusion} described by the stochastic
differential equation $\id Y_t=-\theta Y_t \id t +\sigma \id B_t$,
where $(B_t)_{t\in \mathbb R}$ is a Brownian motion, and $\theta \in
[0,\infty)$ and $\sigma >0$ are parameters known as the \emph{damping} and \emph{volatility} respectively.
Norris \cite{Norris88} found an exact solution to the forward problem
when $\theta=0$, obtaining $R^Y$ when $Y_t =\sigma B_t$. This is
referred to as the case of ``Brownian slope''. 
\section{Brownian slope}
\label{Brownian slope}
The result from \cite{Norris88}, that $R^Y$ is an Ito diffusion when
$Y_t=\sigma B_t$ is stated briefly below: 
 
The quantity
\begin{equation}
L_t^x:=\int_0^t \id s\,\delta(X_s-x)
\end{equation}
known as the \emph{occupation density} or \emph{local time}, measures infinitesimally the ``time spent at $x$ by $X$ before $t$''.
Set $X_t=\sigma B_t-t$, where $(B_t)_{t\geq 0}$ is a Brownian motion, started at zero. 
Also, let $T$ denote the first time that $X$ hits level $K< 0$.
It follows from the Ray-Knight theorem on Brownian local time, that the process $(L_T^x)_{x\geq K}$ has generator
\begin{equation}\label{eq:First}
\mathscr{L}=\frac{2}{\sigma^2}\left[ l\left (\frac{\id}{\id l}\right )^2
+ (\mathbf{1}_{\{x\leq 0\}}-l )\frac{\id}{\id l}\right ].
\end{equation}
Since $X_t\rightarrow-\infty$ as $t\rightarrow\infty$, $(X_t)_{t\geq 0}$ has a \emph{final local time}
$F_x:= \lim_{t\rightarrow\infty}L_t^x<\infty$. Taking $K\rightarrow-\infty$, $(F_x)_{x\in\mathbb{R}}$ is a diffusion, and 
$R_a^{\sigma B}=\lim_{b\rightarrow \infty}F_{-a-b}$ is a diffusion on $a\in\mathbb{R}$, with generator
\begin{equation}\label{eq:Second}
\mathscr{L}=\frac{2}{\sigma^2}\left[r\left(\frac{\id }{\id r}\right)^2
+(1-r)\frac{\id}{\id r}\right].
\end{equation}

\begin{figure}[h]
\psfrag{A}[c]{{\small $a$}}
\psfrag{B}[c]{{\small $\sigma$}}
 \includegraphics[width=1.0\linewidth]{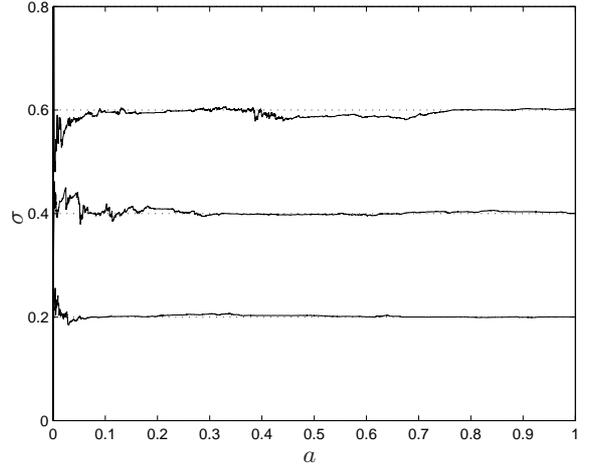}
\caption{Three values of $\sigma$ computed from \eqref{eq:SolutionBM}
  using \eqref{eq:nrw1} for Brownian slope. Note that the rate of convergence is only
  limited by the finiteness of $\Delta t$ and $\Delta a$. The
  Brownian motion is simulated exactly with $\Delta a=2\times 10^{-4}$
  and $\Delta t=10^{-8}$.}
\label{fig2}
\end{figure}

Now consider the problem of finding $\sigma$. Calling our probability measure $\mathbb{P}$, 
it follows immediately from \eqref{eq:First} that
$L_T$ satisfies
\begin{equation}\label{eq:Inverse}
\sigma^2=4\int_J^I\id x\,L_T^x/\langle L \rangle_{J,I}\quad\mathbb{P}\mbox{-a.s.}
\end{equation}
where $I>J>K$, and $\langle L\rangle_{J,I}$ denotes the quadratic variation of $L_T$ on $[J,I]$. Also,
by sending $K\rightarrow-\infty$ in \eqref{eq:Inverse}, or from \eqref{eq:Second}, $R^{\sigma B}$ satisfies
\begin{equation}\label{eq:Solution}
\sigma^2=4\int_I^J\id a\,R_a/\langle R \rangle_{I,J}\quad\mathbb{P}\mbox{-a.s.}
\end{equation}
for $I<J$. At first glance, the expression \eqref{eq:Solution} might 
appear to be the end of the matter, but there is an important sense
in which it is unphysical. A density is of course an
idealization. In practice, one seeks an approximation by measuring the
energy flux over some finite area. For a general density
$(Z_x)_{x\in\mathbb R}$ define the ``discrete local average''
$(\bar Z_n)_{n\in \mathbb Z}$ by
\begin{equation}
\bar Z_n:=\frac{1}{\Delta a}\int_{A_n}\id x\,Z_x
\end{equation}
where $\Delta a>0$ is the width over which the average is taken. The quantity $\bar Z$ is
observable, and we assume that measurements of $\bar Z$ are available. Consider the case
$Z=B$, a Brownian motion. One quickly finds that
\begin{equation}\label{eq:QVDLA}
\mathbb E[(\bar B_{n+1}-\bar B_n)^2]=2\Delta a/3.
\end{equation}
Define a
partition of $[I,J]$ as $\{a_n;n=n_0,...,\tiny{N}\}$, and the
quadratic variation of $\bar Z$ as 
\begin{equation}
\langle \bar Z\rangle_{I,J}:=\lim_{\Delta a\rightarrow 0}Q_{I,J}(\bar
Z)
\end{equation}
where
\begin{equation}
Q_{I,J}(\bar Z):=\sum_{n=n_0}^{N-1}(\bar Z_{n+1}-\bar Z_n)^2.
\end{equation}
It follows from \eqref{eq:QVDLA} that $\langle\bar B\rangle_{I,J}\neq\langle B\rangle_{I,J}$. 
If $\langle \bar B\rangle_{I,J}$ is deterministic \eqref{eq:QVDLA}
implies that 
$\langle \bar B\rangle_{I,J}=2(J-I)/3$ $\mathbb P$-a.s..
This follows from the second moment method if
\begin{align}
&\lim_{\Delta a\rightarrow 0}\mathbb
E[Q_{I,J}^2(\bar B)]\label{eq:bigsec}\\
=&\lim_{\Delta a\rightarrow 0}\mathbb E[Q_{I,J}(\bar B)]^2=4(J-I)^2/9.\label{eq:bigbob}
\end{align}
The expectation in \eqref{eq:bigsec} is not as easy to
evaluate as in the Brownian case (without the local average), since the increments $\bar
B_{n+1}-\bar B_n$ are not 
independent. However, the (quadruple) integrals that result from
multiplying out the squared sum are trivial to compute 
using the correlation function $\mathbb E[B_sB_tB_uB_v]=s(2t+u)$ where
$0\leq s\leq t\leq u\leq v$. For example, for $n_0=0$, $N=1$ we have
\begin{align}
&\mathbb E[Q^2(\bar B)]=\mathbb E\left[(\bar
    B_{1}-\bar B_0)^4\right]\nonumber\\
=&\mathbb E\left[\bar B_0^4-2\bar
B_0^3\bar B_1+6\bar B_0^2\bar B_1^2
-2\bar B_0\bar B_1^3+\bar B_1^4\right].\label{eq:firstsum}
\end{align}
Each of the terms in \eqref{eq:firstsum} can be evaluated by
exchanging the order of the expectation and the integral, and time
ordering. For example, the first term:
\begin{align}
\mathbb E\left[\bar B_0^4\right]=&\frac{4!}{(\Delta
  a)^4}\int_0^{\Delta a} \id v\int_0^v \id u \int_0^u
\id t\int_0^t \id s\,\mathbb E[B_sB_tB_uB_v]\nonumber\\
=&(\Delta a)^2/3.
\end{align}
For larger values of $N$ one can use a computer to sum the terms and
verify that $\mathbb E[Q_{I,J}^2(\bar B)]$ does indeed appear to tend to the
required limit \eqref{eq:bigbob}. Taking
this as an assumption, and
applying the same reasoning to the process defined by
\eqref{eq:Second}, it follows from \eqref{eq:Solution}, in the limit
$\Delta a\rightarrow 0$, that $\bar R^{\sigma B}$ satisfies
\begin{equation}\label{eq:SolutionBM}
\sigma^2=\frac{8}{3}\int_I^J \id a\,\bar R_a/\langle \bar R \rangle_{I,J}\quad\mathbb{P}\mbox{-a.s.}
\end{equation}
where $\bar R_a^Y:= \lim_{\Delta a\rightarrow 0}\bar R_n^Y=R_a^Y$ for
$a\in A_n$. Hence \eqref{eq:SolutionBM} solves the inverse problem for the
Brownian slope. Noting that $\bar R_n^Y=\lim_{\Delta t\rightarrow
  0}\mathbf R_n^Y(\Delta t,\Delta a)$, $\bar R$ can be simulated
using \eqref{eq:nrw1}. 
Fig. \ref{fig2} shows values of $\sigma$ computed using \eqref{eq:SolutionBM}.

The
Brownian slope
model is somewhat unsatisfactory since Brownian motion is
non-stationary and $\lim_{t\rightarrow\infty}|B_t|=\infty$. Of greater
verisimilitude is the model resulting when
$\theta>0$, then $Y$ is a stationary
Ornstein-Uhlenbeck (OU) process, a Gaussian Markov process with
exponential autocorrelation function. This is referred to as the
case of ``OU slope'' and is considered next. 
\section{OU slope}
\label{OU slope}
\begin{figure}[h]
\psfrag{A}[c]{{\small $a$}}
\psfrag{B}[c]{{\small $\sigma$}}
 \includegraphics[width=1.0\linewidth]{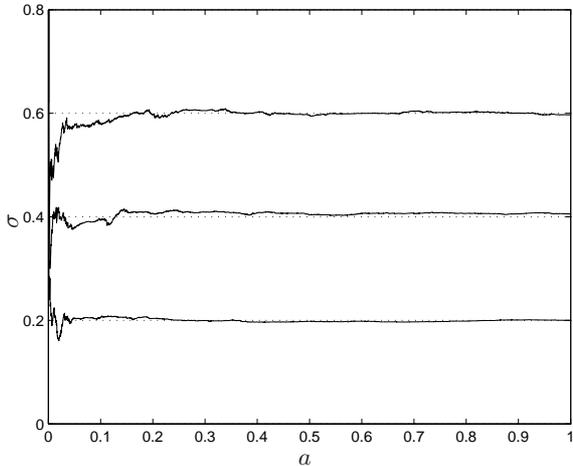}
\caption{Three values of $\sigma$ computed from \eqref{eq:SolutionOU}
  using \eqref{eq:nrw1} for OU slope. As in the Brownian case, the rate of convergence is only
  limited by the finiteness of $\Delta t$ and $\Delta a$. The
  OU process is simulated exactly with $\Delta a=2\times 10^{-4}$
  and $\Delta t=10^{-8}$.}
\label{fig3}
\end{figure}

Define a probability measure
$\mathbb{Q}$ such that $\id\mathbb{Q}:=M_T \id\mathbb{P}$. If $(M_t)_{0\leq
  t \leq T} $ is a martingale then $\mathbb P$ and $\mathbb Q$ are
\emph{equivalent} which means that any
event that has probability one w.r.t. $\mathbb P$ must also have
probability one w.r.t. $\mathbb Q$.
Let $(B_t)_{t\geq 0}$ be a $\mathbb{P}$-Brownian motion, started at zero. Then $(M_t)_{0\leq t \leq T} $ where
\begin{align}\label{eq:POD}
&M_t=\\
&\mathrm{exp}\left(-\theta\int_0^t\mathbf{1}_{ \{s \leq -K \}}B_s \id B_s-
\frac{\theta^2}{2}\int_0^t \id s\,\mathbf{1}_{\{s \leq
  -K\}}B_s^2\right)\nonumber
\end{align}
is a martingale, which can be verified, for example, with the help of
Example 3, p.233, \cite{Liptser01}.
Now, for the process $(X_t)_{0\leq t\leq T}$ defined as above, it
follows from the Girsanov theorem (see for example \cite{Rogers00}) that
\begin{equation}\label{eq:OUchange}
\id X_t=-\mathbf{1}_{\{t \leq -K\}}\theta(X_t+ t)\id t-\id t+\sigma \id W_t\quad\mbox{w.r.t.}\,\,\mathbb{Q}
\end{equation}
where $(W_t)_{0\leq t \leq T} $ is a $\mathbb{Q}$-Brownian motion. The process $(U_t)_{0\leq t\leq -K\wedge T}$ where
$U_t:= X_t+t$
is a $\large\mathbb{Q}$-OU process, such that
$\id U_t=-\theta U_t \id t+\sigma \id W_t$ w.r.t. $\mathbb{Q}$.
By the equivalence of $\mathbb{P}$ and $\mathbb{Q}$, \eqref{eq:Inverse} implies that
\begin{equation}
\sigma^2=4\int_J^I \id x\,L_T^x/\langle L \rangle_{J,I}
\quad\mathbb{Q}\mbox{-a.s.} \label{eq:Gfunctional} 
\end{equation}
where $I>J>K$. Note that $L_T$ is \emph{not} Markov under $\mathbb Q$. The remaining steps are understood to be w.r.t. $\mathbb{Q}$. 
Taking $K\rightarrow-\infty$, $(X_t)_{t\geq 0}$ has a final local time $F_x<\infty$. Then,
$R_a^U=\lim_{b\rightarrow\infty}F_{-a-b}$ where $(U_t)_{t\in\mathbb{R}}$ is a stationary OU process described by
$\id U_t=-\theta U_t \id t+\sigma \id W_t$,
and $(W_t)_{t\in\mathbb{R}}$ is a Brownian motion. It now follows from \eqref{eq:Gfunctional}, 
in this limit, that  $R^U$ satisfies
\begin{equation}\label{eq:Solution2}
\sigma^2=4\int_I^J \id a\,R_a/\langle R \rangle_{I,J}\quad\mathbb{Q}\mbox{-a.s.}
\end{equation} 
for $I<J$. Then, noting that the change-of-measure argument goes
through as above, and assuming \eqref{eq:bigbob}, $\bar R^U$ satisfies
\begin{equation}\label{eq:SolutionOU}
\sigma^2=\frac{8}{3}\int_I^J \id a\,\bar R_a/\langle \bar R \rangle_{I,J}\quad\mathbb{Q}\mbox{-a.s.}
\end{equation}
where $\bar R_a$ is defined as in \eqref{eq:SolutionBM}. Hence
$\sigma$ can be obtained given the ray-density on any
non-zero interval in the case of OU slope, see Fig. \ref{fig3}. 
We emphasize that the
expression is valid for arbitrary $\theta\in [0,\infty)$. 

\begin{figure}[h]
 \includegraphics[width=1.0\linewidth]{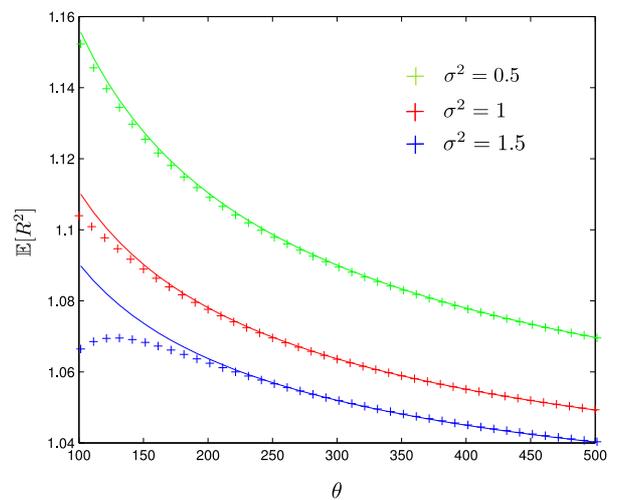}
\caption{(color online) An approximate expression for the second moment. Solid lines are
  \eqref{eq:approx}. The `$+$' denote exact values from numerical
  integration of \eqref{eq:mommeth}.}
\label{fig4}
\end{figure}

With regard to $\theta$, consider the moments of $\bar R^U$. It is not
hard to see that the moments of $\bar R^U$ are equal to the moments of
$R^U$ in the limit
$\Delta a \rightarrow 0$. Therefore,
for ease of calculation, the moments of $R^U$ are considered, and
the superscript is dropped to aid clarity. One quickly
finds that $\mathbb E[R]=1$. For
the second moment, define the process $\{R_a(t_0);a\in\mathbb{R}\}$ as
\begin{equation}\label{eq:One}
R_a(t_0):=\int_{t_0}^\infty \id t\,\delta(U_t-t+a).
\end{equation}
where $U$ is an OU process with
transition density
\begin{align}
&\rho(U_t=y|U_s=x)=\\
&\frac{\sqrt{\theta}}{[\pi \sigma^2(1-e^{-2\theta |t-s|})]^{1/2}}\mathrm{exp}\left[-\frac{\theta(y-xe^{-\theta |t-s|})^2}{\sigma^2(1-e^{-2\theta |t-s|})}\right]\nonumber
\end{align}
see for example \cite{handbook}. For $X_t:= U_t-t$, it follows that
\begin{align}
&\mathbb E[R_a^2(t_0)|\{X_{t_0}=x_0\}]\\
=&2\int_{-a}^\infty \id q\,\rho(U_{q+a}=q+t_0|U_0=u_0)\mathbb E[R_q(t_0)|\{X_{t_0}=q\}].\nonumber
\end{align}
The unconditional second moment can be found by taking
$t_0\rightarrow -\infty$, or, equivalently, by fixing $t_0=0$ and
sending $a\rightarrow\infty$:
\begin{align}
&\mathbb E[R^2]=\lim_{a\rightarrow\infty}\mathbb
E[R_a^2(0)|\{X_{0}=q\}]\nonumber\\
=&2\int_{\mathbb R}\id q\,\rho(U=q)\mathbb E[R_q(0)|\{X_{0}=q\}]\label{eq:mommeth}
\end{align}
where $\rho(U=q)$ is the point-density of the long-term limit of $U$. This
calculation is similar to the Kac moment formula \cite{handbook}. In general \eqref{eq:mommeth} must be evaluated
numerically. An approximate analytical result is available when the
correlation length of $U$ is small compared to the height of the point
of observation above the
surface, that is $1/\theta \ll 1$. Assume for simplicity that $\sigma^2\sim 1$, then
\begin{equation}\label{eq:approx}
\lim_{\Delta a\rightarrow 0}\mathbb E[\bar R^2]=\mathbb E[R^2]\simeq 1+\frac{1.1}{\sqrt{\theta\sigma^2}}
\end{equation}
holds to a close approximation for $\theta \gtrsim
10^2$, see Fig. \ref{fig4}. Finally, \eqref{eq:approx} combined with \eqref{eq:SolutionOU}
provides a way to estimate of $\theta$ given $\sigma$.
\section{Application to sensing of sea ice thickness and discussion}
\label{ApSum}
An example of an application is the following: The first convincing evidence for
the thinning of the Arctic sea ice was collected using submarine
mounted sonar \cite{wadhams1}. Sonar methods continue to this day to provide
the most reliable source of information on sea ice thickness on large
scales \cite{rothrock}. The thickness of the ice
can be inferred from the range of the mean
surface of the underside. This surface is known to possess fractal
properties, and to have an approximately exponential correlation
function \cite{wadhams2}. The statistics
of the ice surface are important parameters for determining the
relation between the range and the time elapsed before the earliest
return \cite{fuks}. Typically, returns from various reflecting
points are resolvable in time. It follows that the
individual intensities can be added, the sum being proportional to the
ray-density, assuming the geometrical optics approximation. Taking the OU
process as a one-dimensional model of the slope of the surface, estimates of $\{\sigma,\theta\}$ can be obtained given $\bar
R^U$ as outlined above.

To summarize, the example provided exploits fluctuations in the ray-density,
which is proportional to the intensity in incoherent
configurations. The potential scope of the method
is clearly far wider than the example given. As already mentioned, the ray-density features in expressions for the
moments of the intensity in coherent configurations, in the short wavelength
limit. Furthermore,
it is $R$ which appears to dictate the structure of intensity fluctuations on a
small-scale when the outer scale is small \cite{jakeman82}, (for OU
slope this means that $1/\theta \ll 1$).
For finite wavelength, the requirement that the scattered amplitude must satisfy the Helmholtz equation
means that the field cannot be ``nowhere
differentiable'', and will therefore have
zero quadratic variation. However, it is only necessary for fractal
structure to be present down to the resolution of the
measurements.
\section*{Acknowledgments}
The author would like to thank NERC for financial support and Joao Rodrigues for proofreading this letter.





\bibliographystyle{model1a-num-names}
\bibliography{trial.bib}







\end{document}